\documentclass[prl,showpacs,showkeys,nofootinbib,twocolumn,floatfix,
superscriptaddress]{revtex4-1}

\usepackage{graphicx}  %
\usepackage{amsmath}
\usepackage{bm}  %
\usepackage{ulem}

\newcommand{\bea}{\begin{eqnarray}}
\newcommand{\eea}{\end{eqnarray}}
\newcommand{\beq}{\begin{equation}}
\newcommand{\eeq}{\end{equation}}
\newcommand{\bqa}{\begin{eqnarray}}
\newcommand{\eqa}{\end{eqnarray}}

\def\mqo2{{\!\!\!}}

\begin{document}

\preprint{MIT-CTP 4498}

\title{
Two-body and Three-body Contacts \\ 
for Identical Bosons near Unitarity}

\author{D.~Hudson Smith}
\email{smith.7991@osu.edu}
\affiliation{Department of Physics,
         The Ohio State University, Columbus, OH\ 43210, USA\\}

\author{Eric Braaten}
\email{braaten@mps.ohio-state.edu}
\affiliation{Department of Physics,
         The Ohio State University, Columbus, OH\ 43210, USA\\}

\author{Daekyoung Kang}
\email{kang1@mit.edu}
\affiliation{Center for Theoretical Physics, 
Massachusetts Institute of Technology, Cambridge, MA\ 02139,USA\\}

\author{Lucas Platter}
\email{lplatter@gmail.com}
\affiliation{Physics Division, 
Argonne National Laboratory, Argonne, IL\ 60439, USA\\}
\affiliation{Department of Fundamental Physics, Chalmers
  University of Technology, SE-412 96 G\"oteborg, Sweden\\}

\date{\today}

\begin{abstract}
In a recent experiment with ultracold trapped $^{85}$Rb atoms,
Makotyn {\it et al.}\ have studied a quantum-degenerate Bose gas 
in the unitary limit where its scattering length is infinitely large.
We show that the observed momentum distributions 
are compatible with a universal relation 
that expresses the high-momentum tail in terms of
the 2-body contact $C_2$ and the 3-body contact $C_3$.
We determine the contact densities for the unitary Bose gas 
with number density $n$ to be 
${\cal C}_2 \approx 20~n^{4/3}$ and ${\cal C}_3 \approx 2~n^{5/3}$.
We also show that the observed atom loss rate is compatible with that from
3-atom inelastic collisions, which gives a contribution proportional 
to $C_3$, but the loss rate is not compatible with that from
2-atom inelastic collisions, which gives a contribution proportional 
to $C_2$.  We point out that the contacts $C_2$ and $C_3$
could be measured independently by using the virial theorem 
near and at unitarity, respectively.
\end{abstract}

\smallskip
\pacs{31.15.-p,34.50.-s, 67.85.Lm,03.75.Nt,03.75.Ss}
\keywords{
Bose gases, 
scattering of atoms and molecules, operator product expansion. }
\maketitle

{\bf Introduction}.
Ultracold atoms allow the study of many-body systems with simple 
zero-range interactions whose strength, which is given by the 
S-wave scattering length $a$, can be controlled experimentally.
These studies are directly relevant to problems in other areas
of physics in which an accidental fine tuning makes $a$ 
much larger than the range of interactions.
In particular, it is relevant to nuclear physics, because nucleons 
have relatively large scattering lengths and because the parameters
of QCD are near critical values for which those scattering lengths 
are infinite \cite{Braaten:2003eu}.
In the {\it unitary limit} where $a$ is infinitely large,
it no longer provides a length scale.
One might therefore expect the interactions to be scale invariant,
so that the only length scales are provided by environmental parameters,
such as the temperature $T$ and the number  density $n$.
This expectation is realized in the simplest Fermi gas, 
which consists of fermions with two spin states.
There have been extensive studies, both experimental and theoretical, 
of the unitary Fermi gas \cite{IKS:0801}.

The simplest Bose gas consists of identical bosons.
The unitary Bose gas is qualitatively different from the simplest 
unitary Fermi gas in two important ways.  The obvious difference comes
from the statistics of the particles.
The other important qualitative difference is that scale invariance 
in the unitary Bose gas is broken by the {\it Efimov effect},
which is the existence of infinitely many 
3-body bound states ({\it Efimov trimers}) whose binding energies
differ by powers of $e^{2\pi/s_0} \approx 515$,
where  $s_0 \approx 1.00624$ \cite{Efimov70}.
This difference is shared with more complicated Fermi gases,
including fermions with three spin states and nucleons near the 
QCD critical point for infinite nucleon scattering lengths.
The breaking of scale invariance by Efimov physics 
introduces a length scale $1/\kappa_*$, where $\kappa_*$
is the binding momentum of one of the Efimov trimers at unitarity,
but physical observables can only depend log-periodically on $\kappa_*$
\cite{Braaten:2004rn}.
This  {\it anomalous symmetry breaking}
can give rise to logarithmic scaling violations at unitarity.
 
Experimental studies of the unitary Bose gas using ultracold atoms
have been hindered by atom losses from inelastic collisions.
In the low-density limit, the rate of decrease in the number density $n$
from 3-body recombination into a deeply bound diatomic molecule ({\it deep dimer})
is proportional to $a^4 n^3$, 
so it grows dramatically as $a$ is increased.
If there was a well-defined unitary limit in which $n$ provided the only length scale,
$dn/dt$ would be proportional to $n^{5/3}$.
The plausibility of a well-defined unitary limit was increased by experimental studies 
of dilute thermal gases of $^7$Li atoms \cite{EcoleNormale:1212}
and of $^{39}$K atoms \cite{Cambridge:1307} 
and by exact theoretical calculations of the loss rate 
for a dilute thermal Bose gas \cite{EcoleNormale:1212}, 
all of which showed that $dn/dt$ at unitarity is proportional to $n^3/T^2$. 
Recently Makotyn {\it et al.}\ have carried out the first studies of 
a quantum-degenerate Bose gas  at unitarity using $^{85}$Rb atoms \cite{JILA:1308}.
They found that, after a quick ramp of a Bose-Einstein condensate (BEC) 
to unitarity, the time scale for the saturation of the momentum distribution 
was significantly shorter than the time scale for atom loss.

Theoretical studies of the unitary Bose gas have been hindered by the
absence of rigorous  theoretical methods that can be used to calculate its properties
with controlled errors. Theoretical studies of the unitary Fermi gas have
faced similar problems, but the absolute stability of 
the system allows the use of Monte Carlo methods that have 
controlled errors.  In the case of the unitary Bose gas,
the possibility of recombination into deeply bound Efimov trimers guarantees 
that, even in the absence of inelastic collisions, the system can be at best metastable.

An alternative to directly calculating the properties of a many-body system
is to use exact solutions to few-body problems
to derive {\it universal relations} between various properties of the system
that  hold for all possible states.
Universal relations for fermions with two spin states were first derived by 
Shina Tan \cite{Tan0505,Tan0508,Tan0803}.  They all involve the {\it 2-body contact} $C_2$.
It is an extensive quantity that can be expressed as 
the integral over space of the  {\it 2-body contact density} ${\cal C}_2$,
which has dimensions (length)$^{-4}$
and can be interpreted as the number of pairs per (volume)$^{4/3}$.
The 2-body contact plays an important role in many of the most important
probes of ultracold fermionic atoms \cite{Braaten:2010if}.
Universal relations for identical bosons were first derived by 
Braaten, Kang, and Platter \cite{Braaten:2011sz}.  They involve not only 
$C_2$ but also the  {\it 3-body contact} $C_3$.
It is an extensive quantity that can be expressed as 
the integral over space of the  {\it 3-body contact density} ${\cal C}_3$,
which has dimensions (length)$^{-5}$ and can be interpreted 
as the number of triples per (volume)$^{5/3}$.

In this Letter, we present universal relations for the 
loss rate of a Bose gas from inelastic 2-atom and 3-atom 
collisions.  We show that the momentum distributions at 
unitarity in the JILA experiment of  Ref.~\cite{JILA:1308} are consistent 
with the universal relation for the tail of the momentum 
distribution in Ref.~\cite{Braaten:2011sz}, and can be used to determine 
${\cal C}_2$ and ${\cal C}_3$ for the unitary Bose gas.  
The result for  ${\cal C}_3$ is 
consistent with the atom loss rate in the JILA experiment 
being dominated by 3-atom inelastic collisions.
In our analysis, we assume that the unitary Bose gase in the 
JILA experiment is in a locally equilibrated metastable state, 
and we ignore the possibility that transient or turbulent phenomena 
could produce steady-state momentum distributions.

{\bf Contacts for identical bosons}.
The 2-body contact $C_2$ and the 3-body contact $C_3$ for a 
state with energy $E$ can be defined
in terms of derivatives of 
$E$ at fixed entropy \cite{Braaten:2011sz}: 
\begin{subequations}
\begin{eqnarray}
\left( a \frac{\partial E}{ \partial a} \right)_{\!\! \kappa_*} &=& 
\frac{\hbar^2}{8\pi m a} ~ C_2,
\label{E-C2}
\\
\left( \kappa_* \frac{\partial E}{\partial \kappa_*} \right)_{\!\! a} &=& 
- \frac{2 \hbar^2}{m} ~ C_3.
\label{E-C3}
\end{eqnarray}
\label{E-C2,3}%
\end{subequations}
Eq.~(\ref{E-C2}) can be used as an operational definition of $C_2$
if the scattering length $a$ can be controlled experimentally.
The normalization of $C_2$ has been chosen so that the tail of the momentum
distribution at large wavenumber $k$ (given below in Eq.~(\ref{n-k}))
is $C_2/k^4$.  
The normalization of $C_3$ in Eq.~(\ref{E-C3}) implies that the 
3-body contact in the unitary limit for an Efimov trimer 
with binding energy $\hbar^2 \kappa_*^2/m$ is $\kappa_*^2$.  
The value of $\kappa_*$ can be inferred from the 
scattering length $a_-$ at which that Efimov trimer 
crosses the 3-atom threshold, producing a  resonance 
in the 3-body recombination rate.  They are related by 
a universal constant: $a_- \kappa_* = -1.50763$ \cite{GME:0802} . 
In the case of $^{85}$Rb atoms, a 3-body recombination resonance 
was observed by Wild {\it et al.}\ at
$a_- = -759(6)~a_0$ with inelasticity parameter 
$\eta_* = 0.057(2)$ \cite{JILA:1112}.

The contacts $C_2$ and $C_3$ determine the high-momentum tail
in the momentum distribution $n(k)$.
We normalize $n(k)$ so that 
the total number of atoms is $N = \int d^3k~n(k)/(2 \pi)^3$.
A systematic expansion for $n(k)$ 
at large wavenumber $k$ can be derived using 
the operator product expansion (OPE) for the quantum field operators
$\psi$ and $\psi^\dagger$ \cite{Braaten:2008uh}.
The universal relation for the 
tail of the momentum distribution for identical bosons 
was derived in Ref.~\cite{Braaten:2011sz}:
\begin{eqnarray}
k^4 n(k) \longrightarrow C_2
+ \frac{A \sin[2 s_0 \ln(k/\kappa_*)+ \phi]}{k} C_3 + \ldots ,
\label{n-k}
\end{eqnarray}
where $A = 89.2626$ and $\phi = -1.33813$. 
The additional terms are suppressed by higher powers of $1/k$ 
that may be noninteger.

{\bf Inelastic loss rates}.
One complication of $^{85}$Rb atoms is that the only hyperfine state
with a Feshbach resonance that can be used to control 
the scattering length has a two-atom inelastic scattering channel
into a pair of atoms in a lower hyperfine state.
The scattering length $a$ is therefore complex 
with a negative imaginary part.  The imaginary part of $1/a$
is essentially constant, independent of the magnetic field \cite{KTJ:0408}:  
Im$(1/a) = 1/(1.44 \times 10^7~a_0)$.
The 2-atom inelastic scattering channel gives a contribution 
to the loss rate of low-energy atoms that is proportional to the 
2-body contact  \cite{Braaten:2013eya}.
This follows from the fact that the effects of 2-particle inelastic scattering 
with large energy release on a system of low-energy particles 
can be taken into account through an antihermitian term in the Hamiltonian
that allows a pair of particles to disappear if they are sufficiently close together.
In a quantum field theory framework, the antihermitian term 
in the Hamiltonian density can be chosen to be the local operator
$\psi^\dagger \psi^\dagger\psi\psi$ multiplied by an imaginary coefficient.
This same operator multiplied by an appropriate ultraviolet-sensitive coefficient
is the 2-body contact density operator \cite{Braaten:2008uh}. 
The loss rate $dN/dt$ can be expressed as the
double integral over space of a correlator of the number density
$\psi^\dagger \psi$ and the 2-body contact density  \cite{Braaten:2013eya}.
Using the commutation relations for $\psi$,
the loss rate can be expressed in the form
\begin{equation}
\frac{dN}{dt}  = - \frac{\hbar}{2 \pi m} ~
{\rm Im}(1/a)~\left(  C_2 + \ldots \right).
\label{2atomloss}
\end{equation}
The coefficient of $C_2$ is the same as for fermions 
with two spin states in Ref.~\cite{Braaten:2013eya}. 
The additional terms  in Eq.~(\ref{2atomloss}) come from the 
integral of the normal-ordered correlator,
which is zero in a system consisting of fewer than three atoms.
If these terms are suppressed,
the $C_2$ term  in Eq.~(\ref{2atomloss}) alone  
provides a good estimate for the loss rate.

If the effects of two-atom inelastic scattering are negligible,
the dominant mechanism for atom loss should be
3-atom inelastic scattering.
The effects of 3-particle inelastic scattering 
with large energy release on a system of low-energy particles 
can be taken into account through an antihermitian term in the Hamiltonian
that allows three particles to disappear if they are all sufficiently close together.
In a quantum field theory framework, the antihermitian term 
in the Hamiltonian density can be chosen to be the local operator
$\psi^\dagger  \psi^\dagger \psi^\dagger \psi\psi \psi$ 
multiplied by an imaginary coefficient.
This same operator multiplied by an appropriate ultraviolet-sensitive coefficient
is the 3-body contact density operator \cite{Braaten:2011sz}. 
Using the methods of Ref.~\cite{Braaten:2013eya},
$dN/dt$ can be expressed as the
double integral over space of a correlator of 
$\psi^\dagger \psi$ and the 3-body contact density.
Using the commutation relations for $\psi$,
the loss rate can be expressed in the form
\begin{equation}
\frac{dN}{dt}  = - \frac{12 \eta_* \hbar}{s_0 m} ~
\left(  C_3 + \ldots \right).
\label{3atomloss}
\end{equation}
The leading term in the expansion was first given by 
Werner and Castin \cite{WC:1210}.
The additional terms come from the integral of the normal-ordered correlator,
which is zero in a system consisting of fewer than four atoms.
If these terms are suppressed,
the $C_3$ term  in Eq.~(\ref{3atomloss}) alone 
provides a good estimate for the loss rate.

{\bf Contact densities}.
The contacts $C_2$ and $C_3$ for a system of trapped atoms 
can be determined using the local density approximation
if the contact densities ${\cal C}_2$ and ${\cal C}_3$
are known for the corresponding homogeneous system.
The contact densities for a homogeneous dilute Bose-Einstein condensate (BEC)  
at zero temperature can be obtained analytically.
The 2-body contact density can be determined from the operational definition 
in Eq.~(\ref{E-C2}).
The 3-body contact density is most easily determined by matching 
Eq.~(\ref{3atomloss}) for the atom loss rate
with the universal result for the loss rate from 3-body recombination 
into deep dimers \cite{Braaten:2004rn}  in the limit $\eta_* \to 0$.
The additional terms  in Eq.~(\ref{3atomloss}) are
suppressed by powers of $n a^3$.
The contact densities for the dilute BEC  are
\begin{subequations}
\begin{eqnarray}
{\cal C}_2 &=& 16 \pi^2 a^2 n^2,
\label{C2-BEC}
\\
{\cal C}_3 &\approx& 
\frac{16 \pi^2  (4\pi - 3\sqrt{3}) s_0  \cosh(\pi s_0)}
    {3 \sinh^3(\pi s_0)} a^4 n^3.
\label{C3-BEC}
\end{eqnarray}
\label{C2,3-BEC}%
\end{subequations}
In Eq.~(\ref{C3-BEC}), we have neglected log-periodic effects that
are numerically suppressed by powers of $e^{-2 \pi s_0} \approx 1/557$.
In Ref.~\cite{JILA:1112}, Wild {et al.}\  put an upper bound on $C_3$ 
for a dilute BEC of $^{85}$Rb atoms. The 3-body contact obtained
using ${\cal C}_3$ in Eq.~(\ref{C3-BEC}) is a factor of 30 below that upper bound.

The contact densities for a homogeneous dilute thermal Bose gas at unitarity  
can also be obtained analytically.
The 2-body contact density in this limit can be obtained by
adapting the analogous calculation for fermions 
in Ref.~\cite{Braaten:2013eya}.
The 3-body contact density is most easily determined by matching 
Eq.~(\ref{3atomloss}) for the atom loss rate
with the exact universal result for the loss rate from 3-body recombination 
into deep dimers \cite{EcoleNormale:1212} in the limit $\eta_* \to 0$. 
The additional terms in Eq.~(\ref{3atomloss})  are
suppressed by powers of $n \lambda_T ^3$,
where $\lambda_T = (2 \pi \hbar^2/m k_B T)^{1/2}$. 
The contact densities for the dilute thermal gas at unitarity are
\begin{subequations}
\begin{eqnarray}
{\cal C}_2 &=& 32 \pi \lambda_T ^2 n^2,
\label{C2-thermal}
\\
{\cal C}_3 &\approx& 
3\sqrt{3}  s_0 \lambda_T ^4 n^3.
\label{C3-thermal}
\end{eqnarray}
\label{C2,3-thermal}%
\end{subequations}
In Eq.~(\ref{C3-thermal}), we have neglected log-periodic effects that
are numerically suppressed by powers of $e^{- \pi s_0} \approx 1/24$.

Exact results for the contact densities 
at unitarity for a homogeneous quantum-degenerate Bose gas
at zero temperature are not known.
If we assume that log-periodic effects are numerically suppressed,
as they are in Eqs.~(\ref{C3-BEC}) and (\ref{C3-thermal}),
the only important length scale for the homogeneous system
is provided by the number density.
If we assume that the contact densities depend weakly on $\kappa_*$,
they must, by dimensional analysis, have the form
\begin{subequations}
\begin{eqnarray}
{\cal C}_2 &\approx& \alpha n^{4/3},
\label{C2-unitary}
\\
{\cal C}_3 &\approx& \beta n^{5/3},
\label{C3-unitary}
\end{eqnarray}
\label{C2,3-unitary}%
\end{subequations}
where $\alpha$ and $\beta$ are numerical constants.
Some values of $\alpha$ obtained in previous attempts 
to calculate ${\cal C}_2$ for the unitary Bose gas are
10.3 \cite{DvHS:1107}, 32 \cite{vHS:1302}, 160 \cite{YR:1308},
and 12 \cite{SCDKGRHB:1309}.
The values in Refs.~\cite{DvHS:1107,vHS:1302} were calculated
for an equilibrium system, while those in 
Refs.~\cite{YR:1308,SCDKGRHB:1309} were calculated for a system 
quenched to unitarity.  All of these calculations used 
uncontrolled approximations.
The local density approximations for the contacts 
of trapped atoms are $C_2 = \alpha N \langle n^{1/3} \rangle$
and $C_3 = \beta N \langle n^{2/3} \rangle$.

\begin{figure}[t]
\centerline{ \includegraphics*[width=8.6cm,clip=true]{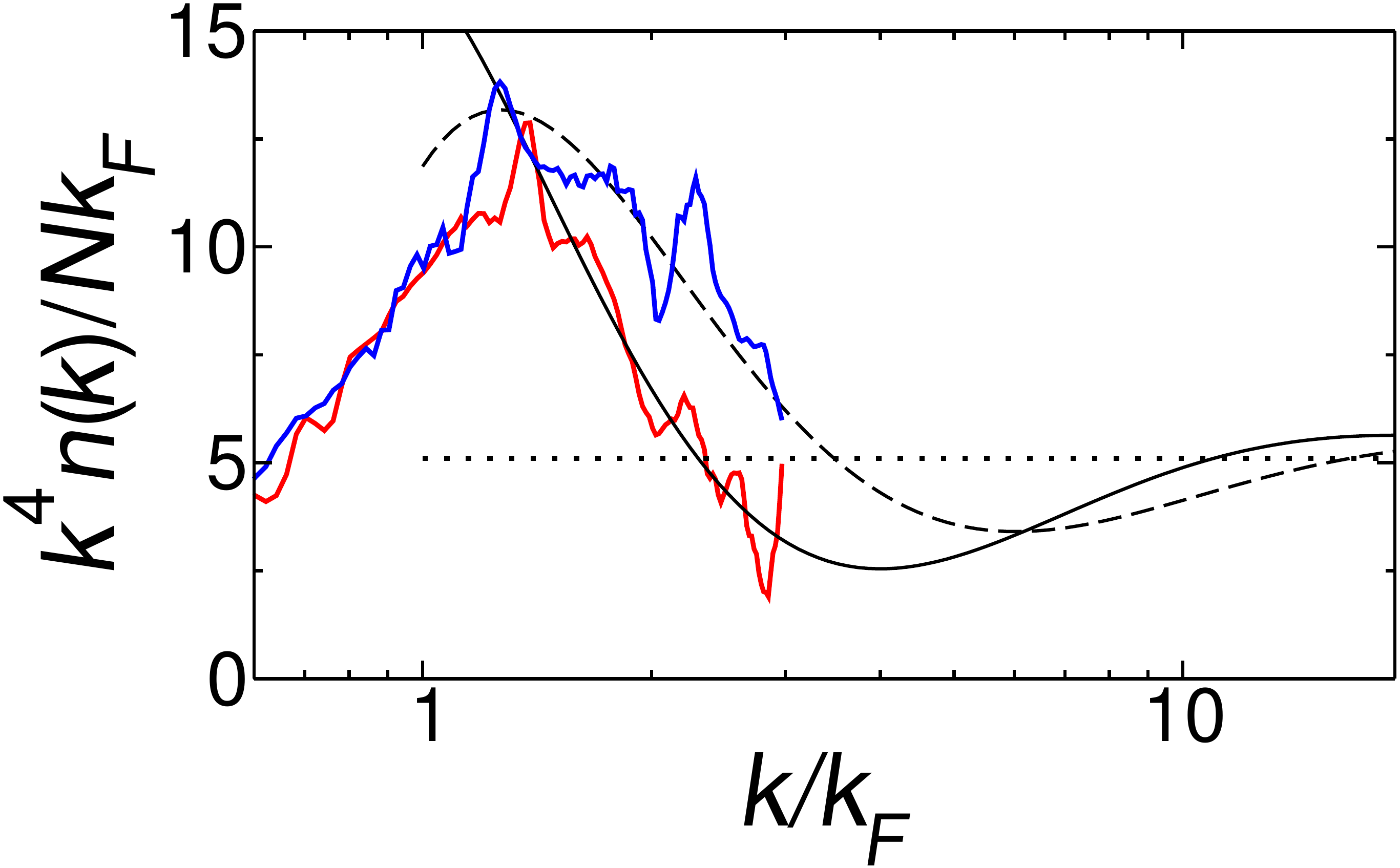} }
\vspace*{0.0cm}
\caption{Momentum distributions for the unitary Bose gas.
The dimensionless quantity $k^4 n(k)/Nk_F$, 
where $k_F = (6 \pi^2 \langle n \rangle)^{1/3}$,
is plotted as a function of $k/k_F$.
The data from the JILA group in  Ref.~\cite{JILA:1308} 
is for two average densities:
$\langle n \rangle = 5.5 \times 10^{12}/{\rm cm}^3$ 
(red line with lower tail) and 
$1.6 \times 10^{12}/{\rm cm}^3$ 
(blue line with higher tail)~\cite{JILA:1308}.
The solid curve through the higher-$\langle n \rangle$ data 
is a 2-parameter fit obtained by adjusting $C_2$ and $C_3$.  
The dashed curve through the lower-$\langle n \rangle$ data 
is a parameter-free prediction obtained by scaling  
$C_2$ and $C_3$ from the higher-$\langle n \rangle$ fit. 
The horizontal dotted line is the contribution to both distributions from $C_2$.}
\label{fig:momtail}
\end{figure}

{\bf Momentum distributions}.
In the experiment of Ref.~\cite{JILA:1308}, a BEC of $^{85}$Rb atoms 
was quickly ramped to unitarity. 
The resulting clouds had approximately Thomas-Fermi
distributions with about 60,000 atoms 
and an average number density $\langle n \rangle$
of either $5.5\times10^{12}/{\rm cm}^3$ or $1.6\times10^{12}/{\rm cm}^3$.
The JILA group measured the momentum distribution $n(k)$  after
a variable holding time at unitarity.  They observed that $n(k)$
saturates in approximately 0.1~ms at the higher density 
and 0.2~ms at the lower density, both of which are
significantly shorter than the atom-loss time scale, 0.6~ms.
The distributions $k^4n(k)$ are plotted in Fig.~\ref{fig:momtail} 
using dimensionless variables obtained by scaling 
by $k_F = (6 \pi^2 \langle n \rangle)^{1/3}$. 
The scaled distributions for the two densities
agree well for $k < 1.1~k_F$, but they differ dramatically for $k > 1.1~k_F$,
indicating large scaling violations in the tails of the momentum
distributions.
According to Eq.~\eqref{n-k}, $k^4n(k)$ should asymptotically approach 
the constant $C_2$ at large $k$, but the distributions in Fig.~\ref{fig:momtail} 
do not appear to be approaching a constant  for either density.

We assume that the data for $k > 1.5~k_F$  in Fig.~\ref{fig:momtail} 
is part of the tail of the momentum distribution that is determined
by $C_2$ and $C_3$ according to Eq.~\eqref{n-k}. 
The positions of the local maxima and minima in the tail 
are predicted in terms of $\kappa_*$, which is determined by the 
Efimov loss resonance observed in Ref.~\cite{JILA:1112}.  
In particular, there should be a minimum at $0.71~\kappa_*$,
which is $3.9~k_F$ for the higher $\langle n\rangle$ 
and $5.8~k_F$ for the lower $\langle n\rangle$.
Fitting Eq.~\eqref{n-k} to the momentum distribution for
$\langle n \rangle = 5.5\times10^{12}/{\rm cm}^3$ from $k=1.5~k_F$ to
$k=3.0~k_F$, we obtain $\alpha=22(1)$ and $\beta=2.1(1)$.
The errors are lower bounds on the uncertainties,
because there are systematic errors in the JILA
experiment that were not quantified. 
The value of $\alpha$
agrees to within a factor of 2 with the previous 
estimates of Refs.~\cite{DvHS:1107,vHS:1302,SCDKGRHB:1309}.
The fitted curve in Fig.~\ref{fig:momtail} predicts that, beyond
the range of the measured data, $k^4 n(k)$ should increase  
and asymptotically approach $C_2$.
Having fit $\alpha$ and $\beta$ to the higher-$\langle n\rangle$ data,
the tail of the momentum distribution for other values of $\langle n\rangle$ can be predicted 
without any adjustable parameters.  The prediction for 
$\langle n \rangle=1.6\times10^{12}/{\rm cm}^3$
is shown in Fig.~\ref{fig:momtail} and is in good agreement with
the data.
Thus the observed scaling violations in the tails of the momentum
distributions are explained by the log-periodic dependence
of the coefficient of the $C_3/k^5$ term in Eq.~\eqref{n-k} on $k/\kappa_*$.

{\bf Atom loss rate}.
The loss of $^{85}$Rb atoms from a trapping potential comes from
inelastic 2-atom collisions,
which gives the $C_2$ term in Eq.~(\ref{2atomloss}),
and from  inelastic 3-atom collisions,
which gives the $C_3$ term in Eq.~(\ref{3atomloss}).
The initial loss rate for trapped atoms determines 
a time constant $\tau$ defined by $dN/dt = - (1/\tau)N$.
In the JILA experiment in Ref.~\cite{JILA:1308},
$\tau$ was determined to be $0.63 \pm 0.03$~ms for 
$\langle n \rangle = 5.5 \times 10^{12}/{\rm cm}^3$.
If we assume the dominant loss mechanism is 
2-atom inelastic collisions as in Eq.~(\ref{2atomloss}) and
use $\tau$ to estimate  $C_2$,
we obtain $\alpha \sim 6000$.
This is more than 30 times larger than any of the 
estimates in Refs.~\cite{DvHS:1107,vHS:1302,YR:1308,SCDKGRHB:1309},
which suggests that 2-atom inelastic collisions
are  unlikely to give a significant contribution to the observed atom losses.
If we assume the dominant loss mechanism is 
3-atom inelastic collisions as in Eq.~(\ref{3atomloss}) 
and use $\tau$ to estimate $C_3$,
we obtain $\beta \sim 1$.
This is within a factor of 2 of the value we
obtained by fitting the momentum distributions.
This makes it plausible that 3-atom inelastic collisions
are the dominant mechanism for the observed atom losses.
The time constant $\tau$ is increased by the suppression factor 
of $\eta_* = 0.06$ in the expression for the loss rate 
in Eq.~(\ref{3atomloss}).

{\bf Other probes of the contacts}.
The virial theorem 
for identical bosons trapped in a harmonic potential 
was first derived by Werner \cite{W:0803}:
\begin{equation}
(T + U) - V  = - \frac{\hbar^2}{16\pi m a} ~ C_2 - \frac{\hbar^2}{m} ~ C_3,
\label{virial}
\end{equation}
where $T$, $U$, and $V$ are the
kinetic, interaction, and potential energies, respectively.
This implies that $C_3$ at unitarity can be determined from
the difference between $T+U$ and $V$
and that $C_2$ can be determined from the slope of that 
difference as a function of $1/a$.
The virial theorem for fermions with two spin states is Eq.~(\ref{virial})
with $C_3=0$.  This universal relation
has been tested by a group at JILA by measuring $T+U$, $V$,
and $C_2$ separately as functions of $a$ for
ultracold trapped $^{40}$K atoms \cite{JILA:1002}.
Similar measurements of $T+U$ and $V$ for identical bosons
near unitarity could be used to determine $C_2$ and $C_3$.

Another way to determine $C_2$ and $C_3$
is using rf spectroscopy, in which a radio-frequency signal 
transfers atoms to a different hyperfine state.
Universal relations for the rf spectroscopy of identical bosons
were derived in Ref.~\cite{Braaten:2011sz}.
They predict scaling violations 
in the high-frequency tail.
 Their observation would add to the compelling theoretical evidence 
from  scaling violations in the tail of the 
momentum distribution that the experiment
in Ref.~\cite{JILA:1308} was studying 
a locally equilibrated metastable state of the unitary Bose gas.

\begin{acknowledgments}
This research was supported in part by the
National Science Foundation under grant PHY-1310862
and by the US Department of Energy, Office of Nuclear Physics, 
under contracts DE-AC02-06CH11357 and DE-FG02-94ER40818.
We thank the JILA group of Eric Cornell and Debbie Jin
for providing the data shown in Fig.~\ref{fig:momtail}.
\end{acknowledgments}

\end{document}